\documentclass[9.5 pt,conference]{IEEEtran}

\IEEEoverridecommandlockouts

\usepackage{bm}
\usepackage{cite}
\usepackage{amsmath,amssymb,amsfonts}
\usepackage{colortbl}
\usepackage{multirow}
\usepackage{booktabs}
\definecolor{LightCyan}{rgb}{0.88,1,1}
\definecolor{LightRed}{rgb}{1,0.88,0.88}
\definecolor{Gray}{rgb}{0.8,0.8,0.8}
\definecolor{LightGray}{rgb}{0.92,0.92,0.92}
\definecolor{LightPurple}{RGB}{226, 225, 254}
\usepackage{url}

\usepackage{algorithmic}
\usepackage{graphicx}
\usepackage{textcomp}
\usepackage{xcolor}
\usepackage[pagebackref,breaklinks,colorlinks]{hyperref}
\def\BibTeX{{\rm B\kern-.05em{\sc i\kern-.025em b}\kern-.08em
    T\kern-.1667em\lower.7ex\hbox{E}\kern-.125emX}}
    
\begin{document}

\newcommand{\cy}[1]{{\color{red}{(CY: #1)}}}
\newcommand{\shaobo}[1]{{\color{cyan}{(SH: #1)}}}

\newcommand{\red}[1]{{\textcolor{red}{#1}}}

\title{CLAP-S: Support Set Based Adaptation for Downstream Fiber-optic Acoustic Recognition}


\author{
    \IEEEauthorblockN{Jingchen Sun\textsuperscript{1,2 $^\dagger$}\thanks{$^\dagger$ This work was conducted during an internship at NEC Labs.}, Shaobo Han\textsuperscript{1}, Wataru Kohno\textsuperscript{1}, Changyou Chen\textsuperscript{2}}
    \IEEEauthorblockA{
        \textsuperscript{1}NEC Laboratories America, Inc, USA
        \textsuperscript{2}University at Buffalo, State University of New York, USA \\
    }
}

\maketitle

\begin{abstract}
 
Contrastive Language-Audio Pretraining (CLAP) models have demonstrated unprecedented performance in various acoustic signal recognition tasks. Fiber-optic-based acoustic recognition is one of the most important downstream tasks and plays a significant role in environmental sensing. Adapting CLAP for fiber-optic acoustic recognition has become an active research area. As a non-conventional acoustic sensor, fiber-optic acoustic recognition presents a challenging, domain-specific, low-shot deployment environment with significant domain shifts due to unique frequency response and noise characteristics. To address these challenges, we propose a support-based adaptation method, CLAP-S, which linearly interpolates a CLAP Adapter with the Support Set, leveraging both implicit knowledge through fine-tuning and explicit knowledge retrieved from memory for cross-domain generalization. Experimental results show that our method delivers competitive performance on both laboratory-recorded fiber-optic ESC-50 datasets and a real-world fiber-optic gunshot-firework dataset. Our research also provides valuable insights for other downstream acoustic recognition tasks. The code and gunshot-firework dataset are available at 
\href{https://github.com/Jingchensun/clap-s}{https://github.com/Jingchensun/clap-s}.
\end{abstract}


\vspace{-1em} 

\section{Introduction}

Distributed acoustic sensing (DAS) \cite{Ip_IEEE2022} is a powerful technology that captures acoustic disturbances by measuring phase changes in the backscattered optical signal caused by vibrations. DAS interrogator connecting to one end of the optical fiber (spanning kilometers)  transforms the cable into equally spaced sensing elements with meter-scale spatial resolution, enabling new fiber-based acoustic applications. In particular, DAS is advantageous for harsh environments such as outdoor or underwater settings, where power supply and data transmission pose challenges. Conventional sound recording device such as microphones is considered as \emph{point sensor}, each of which only monitors one small area. In contrast, DAS can utilize long optical fibers as \emph{linear sensors} for wide-area monitoring over vast distances without the need for numerous individual sensors and battery installation. This technology has been successfully applied to pipeline leak detection \cite{tanimola2009distributed}, rail crack detection \cite{fan2019rail}, drone detection \cite{fang2022drone}, utility pole localization \cite{lu2021automatic, jiang2023utility}, seismic monitoring \cite{fernandez2020distributed}, insect activity monitoring \cite{insect}, whale call detection \cite{bouffaut2022eavesdropping}, and underwater surveillance \cite{lu2021distributed}.  


Despite high prospects in a wide range of industrial applications, developing fiber-optic acoustic recognition systems still faces some challenges:
First, field data collection and annotation are labor-intensive and time-consuming. As a result, obtaining sufficient labeled data from DAS for supervised learning is often more difficult than from microphones, especially for rare classes in the long tail. Second, the characteristics of sensing data are influenced by multiple factors including sensor configuration, propagation media, signal source, and optical factors \cite{zhou2013characteristics, tonami2024low}, causing severe domain gaps. Third, users may be interested in recognizing events of new classes that are unseen during training, which leads to an open-set recognition problem. 

Recently, contrastive language-audio pre-training (CLAP) models \cite{clap1, clap2} has emerged as a new paradigm for learning general-purpose audio representations. CLAP have demonstrated strong zero-shot performance in multiple downstream domains. Encouraged by this success, we explore the possibility of adapting CLAP to the fiber-optic acoustic domain of interest. Due to severe domain shifts, directly using CLAP pre-trained on microphone-recorded audio data to recognize fiber-optic acoustic events results in very low zero-shot classification accuracy, e.g., less than 30\% on a 50-class environmental sound dataset recorded from a fiber coil, as shown in Table \ref{main}. 

Existing approaches, such as Prompt Tuning \cite{coop, audio-free, hanif2024palm, sun2023prompt} and Adapter methods \cite{clip-adapter, tip-adapter, wu2023large, kessler2022adapter}, enable efficient fine-tuning on downstream tasks. Prompt Tuning leverages learnable text prompts to maximize the extraction of implicit knowledge \cite{generalization, mem} from pre-trained models but often struggles to incorporate new knowledge when faced with significant domain gaps. In contrast, Adapter methods employ projection layers to explicitly acquire task-specific knowledge. Methods like Tip-Adapter \cite{tip-adapter} and Treff \cite{treff} introduce a learnable adapter to enhance CLAP models in low-shot scenarios, achieving competitive performance. However, the relative importance of implicit knowledge embedded in pre-trained models versus explicit knowledge obtained through fine-tuning remains unclear.

Is the implicit knowledge from the pre-training model always helpful? In this paper, we address the problem of adapting CLAP models for fiber-optic acoustic recognition as a downstream task. 
We focusing on how the pre-trained knowledge, domain shift, and limited labels will affect our adaptation process. To systematically study this problem, we create a fiber-optic version of the ESC-50 dataset \cite{piczak2015dataset} by replaying and recording it under various data acquisition settings in the lab.  We also consider real-world gunshot vs. firework classification dataset collected from existing telecommunication cables in the field \cite{han2024deep}. This task is particularly challenging due to differences in the recording device, the complex outdoor environment, and the sound characteristics of the unseen and new classes. 

 We evaluate the effectiveness of several model adaptation approaches, including prompt tuning \cite{coop, audio-free}, inserted adapter \cite{clip-adapter}, Tip-Adapter \cite{tip-adapter} and Treff \cite{treff}. Different from existing methods, to maximize the use of labeled data, we propose fine-tuning and augmenting CLAP models with the same support set, which facilitates cross-domain generalization based on both implicit knowledge memorized in model parameters and explicit knowledge stored in the support set. The proposed approach consistently improves performance on both lab-collected dataset and the real-world gunshot-firework event classfication dataset. 

\section{Method}


Given a pre-trained CLAP model and a downstream dataset, we assume $K$-shot $N$-class training samples for fine-tuning. For all $NK$ training audios $X_K$, the audio embeddings are represented as ${F}_{\text{train}} = \operatorname{AudioEncoder}(X_K)$, where ${F}_{\text{train}} \in \mathbb{R}^{NK \times C}$, where $C$ is the hidden dimension of the audio encoder. The label vectors are represented as ${L}_{\text{train}} = \operatorname{OneHot}(L_N)$, where ${L}_{\text{train}} \in \mathbb{R}^{NK \times N}$. The audio embeddings and label vectors form the keys and values of the \textbf{Support Set}, respectively. The Support Set stores all new knowledge extracted from the training set.

For a given test audio sample $x$, the normalized audio embedding $u \in \mathbb{R}^{1 \times C}$ is obtained by feeding the sample into the audio encoder. The audio embedding $u$ serves as a \textbf{query}, while the stored audio embeddings ${F}_{\text{train}}$ serve as \textbf{keys}. Cross-attention \cite{attention} is applied between the query and keys. The attention weights are multiplied by the \textbf{values} ${L}_{\text{train}}^T$ to obtain the similarity-based prediction.

\begin{figure}[bpht!]
  \centering
\vspace{-0.5em} \includegraphics[width=0.5\textwidth]{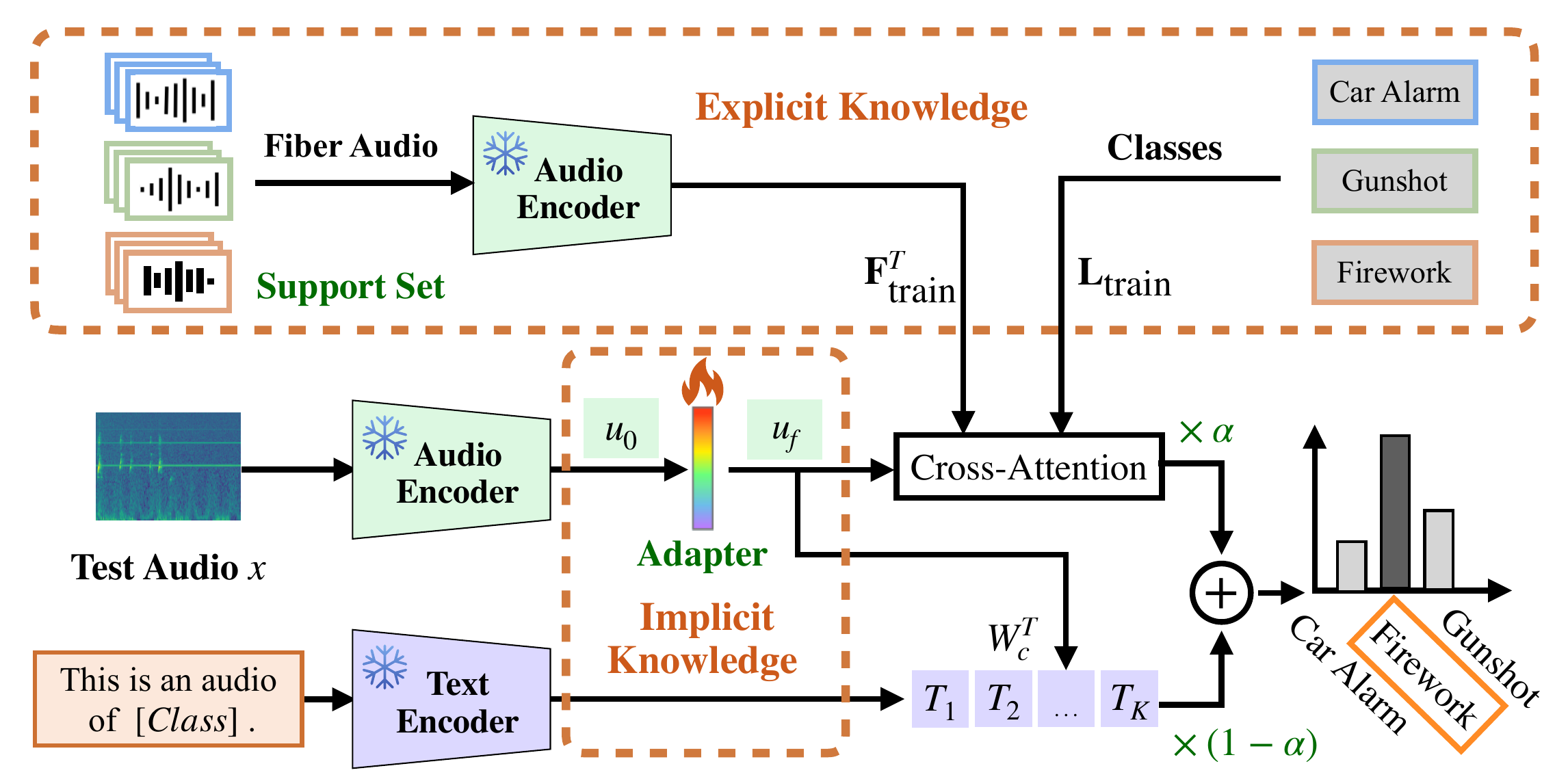}       
  \caption{\textbf{The pipeline of our proposed method.} A test sample is sent to the frozen pre-trained audio encoder and fine-tuned adapter to obtain the embedding, which is then used to perform cross-attention with the keys from the support audio samples. The attention weights are further multiplied by the values of the support set to serve as \textbf{Explicit Knowledge}.  The final prediction is obtained by Linear interpolation with the Explicit Knowledge and the \textbf{Implicit Knowledge} captured by a fine-tuned Adapter .}\label{fig2:pipeline}

\end{figure}

\textbf{Complimentary Mechanisms of  Generalization.} We propose to utilize supervised information in the support set twice, for both the fine-tuning adapter and the key-value support set. Our approach uses a Linear interpolation between both implicit knowledge from model fine-tuning and explicit knowledge from the support set and seeks a balance. The final prediction is derived by 
\begin{align}
p_{\text{final}}(y|x) &= (1-\alpha) p_{\text{clap}}\left(y|x, u\right) + \alpha p_{\text{support}}\left(y|x, u\right)
\label{prediction}
\end{align}
where $p_{\text{clap}}$ is the class distribution from the fine-tuned CLAP model and $p_{\text{support}}$ is the class distribution from the support-set-based class distribution. They are defined as follows:
\begin{align}
p_{\text{clap}}\left(y|x, u\right) &= {u W_c^T}, \nonumber \\
p_{\text{support}}\left(y|x, u\right) &= e^{-\beta\left(1-u F_{\text{train}}^T\right)} L_{\text{train}}^T.  \nonumber
\end{align}
Here, $\alpha \in [0, 1]$ is a tuned parameter balancing the contribution from the implicit knowledge and explicit knowledge from the support set, and $\beta$ is the sharpness parameter. Here $u$ has two different representations, one is the text-aligned audio representation $u_{0}$, and the other one is the task-aligned audio representation $u_{\textrm{f}}$, they obtained by the following equation:
\begin{align}
u_{\textrm{f}} =\operatorname{Adapter}\left(u_{0}\right), \quad {u_{0} = \operatorname{Audio Encoder}(x)}
\end{align}
The $\operatorname{Adapter}$ is a two-layer MLP. $u_{0}$ is obtained by feeding the test audio sample $x$ into the audio encoder, which is aligned with the text embedding during pretraining. And ${u_{\textrm{f}}}$ is derived by passing $u_{0}$ through the two-layer MLP Adapter, while ${u_{\textrm{f}}}$ is aligned with the task during fine-tuning. 

By combining $u_{0}$ and $u_{\textrm{f}}$, we can derive many variants from Equation \ref{prediction}: 
(1) combine $u_{0}$ for $p_{\text{clap}}$ and $u_{\textrm{f}}$ for $p_{\text{support}}$. (2) combine $u_{\textrm{f}}$ for $p_{\text{clap}}$ and $u_{0}$ for $p_{\text{support}}$. (3) combine $u_{\textrm{f}}$ for $p_{\text{clap}}$ and $u_{\textrm{f}}$ for $p_{\text{support}}$. \textbf{These combinations allow us to explore whether text-aligned or task-aligned embeddings are more effective for fiber acoustic recognition.} 
We empirically find that combination (3) proves to be the most effective for our task and we summarize our method two variants:




\textbf{CLAP-S:} Here $u$=$u_{0}$ and  $\alpha = 1$, our final prediction relies solely on generalization through memorization via $p_{\text{support}}$, completely removing the influence of zero-shot or fine-tuning knowledge.

\textbf{CLAP-S$^{+}$: } Here $u$=$u_{\textrm{f}}$ and $0 <\alpha<1$, $p_{\text{clap-s}}$ is obtained by linearly interpolating between the knowledge in $p_{\text{clap}}$ and the explicit knowledge $p_{\text{support}}$ from the support set. This method uses task-aligned embeddings for both classification and retrieval. The pipeline of CLAP-S$^{+}$ is shown in Figure \ref{fig2:pipeline}.

\begin{table}[bpht]
\centering
\caption{Implicit \& Explicit Generalization Combination.}\label{table:clap-s-variants}
\setlength{\tabcolsep}{1pt}
\fontsize{8}{8}\selectfont
\begin{tabular}{@{}ccllc@{}}
\toprule
$p_{\text{clap}}$    & $p_{\text{support}}$   & Combination                    & Method               & ZS \\ \midrule
$u_{0}$     &         & ZS ($\alpha$=0)                & ZS-CLAP              & Yes         \\
        & $u_{0}$     & Support set ($\alpha$=1)               & \textbf{CLAP-S}           & No         \\
$u_{0}$     & $u_{0}$     & ZS + Support set                & Tip-Adapter          & Yes         \\
$u_{0}$     & $u_{\textrm{f}}$     & ZS + Support set$^{\textrm{F}}$   & Tip-Adapter-F & Yes         \\

$u_{\textrm{f}}$ & $u_{\textrm{f}}$     & Adapter +  Support Set$^{+}$    & \textbf{CLAP-S$^+$}       & No         \\ \bottomrule
\end{tabular}
\end{table}

\begin{table*}[htp!]
\centering
\caption{Zero-Shot \& Full-shot Adaption on the Fiber-Optic Acoustic Sensing Dataset.}
\setlength{\tabcolsep}{8pt}
\fontsize{8}{8}\selectfont
\begin{tabular}{@{}lllllllll@{}}
\toprule
                                   & Method                              & Shot & \multicolumn{3}{c}{Laboratory Task} & \multicolumn{2}{c}{Real Task} & Average\\ 
                                   \cmidrule(lr){4-6} \cmidrule(lr){7-8}
                                   &                                     &  & ECM        & FM         & FC        & FMO            & FCO         &   \\
                                   \midrule
\multirow{5}{*}{Trainin-Free}      & {[}class{]}                         & zero & 71.8       & 35.2       & 22.1      & 18.0          & 14.0        & 32.2  \\
                                   & this is {[}class{]}                 & zero & 79.9       & 42.9       & 27.4      & 20.0          & 10.0        & 36.0  \\
                                   & this is an audio of {[}class{]}     & zero & 80.7       & 44.3       & 27.2      & 17.0          & 10.0        & 35.8 \\
                                   & i can hear the sound of {[}class{]} & zero & 77.5       & 40.6       & 27.9      & 16.0          & 16.0         & 35.6  \\

                                   & Tip-Adapter \cite{tip-adapter}                & full & 87.0$_{\pm 0.1}$       & 59.0$_{\pm 0.1}$       & 39.0$_{\pm 0.1}$      & 78.8$_{\pm 1.5}$          & 82.2$_{\pm 1.0}$     & 69.2     \\
                                   & \textbf{CLAP-S (ours)}                & full & 92.0$_{\pm 0.1}$       & 61.4$_{\pm 0.8}$       & 43.0$_{\pm 0.0}$      & 79$_{\pm 0.1}$          & 83$_{\pm 0.1}$     & 71.6     \\\midrule
                                   
                                   \rowcolor{LightGray}
                                   & $\bigtriangleup$      &                              & \textbf{+5.0} $\uparrow$        & \textbf{+2.4} $\uparrow$      & \textbf{+4} $\uparrow$     & \textbf{+0.2} $\uparrow$       & \textbf{+0.8} $\uparrow$    & \textbf{+2.4} $\uparrow$    \\ \midrule
                                   & Prompt Tuning \cite{coop}                      & full & 87.3$_{\pm 1.7}$       & 46.3$_{\pm 1.8}$       & 30.2$_{\pm 1.2}$      & 4.0$_{\pm 4.0}$           & 5.0$_{\pm 5.2}$    & 34.6       \\
                                                                      & Adapter \cite{clip-adapter}                    & full & 92.9$_{\pm 1.1}$        & 68.8$_{\pm 1.0}$        & 48.8$_{\pm 1.7}$       & 81.4$_{\pm 0.9}$           & 90.2$_{\pm 0.7}$    & 76.4        \\
                                    & Treff \cite{treff}               & full & 90.3$_{\pm 0.8}$       & 67.5$_{\pm 0.9}$       & 46.8$_{\pm 1.0}$      & 84.2$_{\pm 1.3}$          & 90.1$_{\pm 1.7}$    & 75.8      \\                                  
\multirow{1}{*}{Training-Required} & Tip-Adapter-F \cite{tip-adapter}               & full & 91.0$_{\pm 0.7}$       & 68.6$_{\pm 1.0}$       & 47.4$_{\pm 0.8}$      & 84.6$_{\pm 1.2}$          & 90.8$_{\pm 0.7}$    & 76.5      \\

                                   & \textbf{CLAP-S$^{+}$ (ours)}                & full & 94.0$_{\pm 1.2}$       & 70.0$_{\pm 0.8}$       & 51.0$_{\pm 1.2}$      & 87.0$_{\pm 1.9}$          & 92.0$_{\pm 1.7}$      & 78.8    \\\midrule
                                   \rowcolor{LightGray}
                                   & $\bigtriangleup$                             &  & \textbf{+2.8 }$\uparrow$       & \textbf{+1.4} $ \uparrow$        & \textbf{+3.0} $\uparrow$      & \textbf{+2.4} $\uparrow$          & \textbf{+1.2} $\uparrow$    & \textbf{+2.3} $\uparrow$      \\ \bottomrule
\end{tabular}
\label{main}
\vspace{-1.3em} 
\end{table*}

\textbf{Relation with Existing works.} The difference between our method and existing method is shown in Table \ref{table:clap-s-variants}.
Compared to Tip-Adapter \cite{tip-adapter} and Treff\cite{treff}, which treats the keys in the whole support as the task-aligned embedding, our CLAP-S$^{+}$ take a more natural approach using embedding from the fine-tuned adapter for both query and key. This leads to stronger generalization in practice (see Table \ref{main} and \ref{ablation}).

\begin{table}[]
\centering
\caption{Dataset Comparison.}
\setlength{\tabcolsep}{2pt}
\fontsize{8}{8}\selectfont
\begin{tabular}{@{}lllllll@{}}
\toprule
                                 &   Dataset Name                        & Classes & Train & Val & Test & Total \\ \midrule
\multirow{3}{*}{Lab Task} & Electric Microphone (ECM) & 50      & 1400  & 400        & 200  & 2000  \\
                                 & Fiber Mandrel (FM)        & 50      & 1400  & 400        & 200  & 2000  \\
                                 & Fiber Coil (FC)        & 50      & 1400  & 400        & 200  & 2000  \\\midrule
\multirow{2}{*}{Real Task}       & Fiber Mandrel Outdoor (FMO)      & 8       & 647   & 167        & 421  & 1235  \\
                                 & Fiber Coil Outdoor (FCO)         & 8       & 406   & 106        & 263  & 775   \\ \bottomrule
\end{tabular}
\label{dataset}
\vspace{-1em} 
\end{table}

\section{Experimental Results}
We use the CLAP model from 2023 as the pre-trained model. 
In all subsequent experiments, each dataset is run five times, and reported the average and standard deviation in the tables. 
We use the AdamW optimizer with a learning rate of 1e-5, batch size of 64, and trained for 20 epochs. All experiments were conducted on an RTX A6000.

\subsection{Datasets}

\noindent\textbf{Laboratory-Recorded Fiber-Optic ESC-50 Dataset} To study the effects of different device domains (and recording environment), we record data using two fiber-optic sensors \cite{text-guide}: (1) a Fiber Mandrel (FM) with a cylinder wrapped in single-mode bare fiber \cite{Kohno:23}, (2) a Fiber Coil (FC), and (3) an Electric Condenser Microphone (ECM8000) as a reference.  We utilize a DAS system to record the ESC-50 dataset \cite{piczak2015dataset}, a public benchmark dataset that includes $2000$ samples from $50$ categories of environment sounds. 




\noindent\textbf{Gunshot-Firework Event Classification Dataset} To further evaluate the ability to distinguish fine-grained event classes and generalize to real-world environments, we consider the outdoor fiber-optic acoustic gunshots-firework event classification  dataset\cite{han2024deep}. It contains $8$ types of real-life sound events: gunshots, crackers, cannons, fountain cannons, high-altitude fireworks, vehicle door slamming, vehicle alarms, and background noise. 
This dataset was collected using DFOS from pre-deployed telecom networks using two types of DAS sensors in the outdoor environment: Fiber Mandrel Outdoor (FMO) and Fiber Coil Outdoor (CO), as two domains with distinct domain shifts. 
The datasets split and comparison are shown in Table \ref{dataset}.


\subsection{Zero-shot \& Full-shot Adaptation }

\noindent\textbf{Zero-shot Adaptation.} This scenario tests the performance of the pre-trained model when we do not have any labeled data for fine-tuning. \textbf{The results show that the domain shift on the real-world task is higher than the laboratory-recorded task.} The laboratory datasets, based on ESC-50, include typical environmental sounds (e.g., dog, engine, rain) frequently found in the pretraining data in the CLAP model under test. In contrast, the real-world dataset features uncommon categories (e.g., `fountain cannon", a label for the firework event) and larger background noise (e.g., wind), which is considered to contribute to poor zero-shot recognition, with accuracy dropping below 20\%. Besides, we also found \textbf{the domain shift in the fiber mandrel is also larger than the fiber coil} as FM achieved a higher accuracy (44.3\%) than the FC (27.9\%), even though they share the same label space.


\noindent\textbf{Full-shot Adaptation.} Once we have enough labeled data, we can inject the new knowledge from these data into the model. If the computing resource is not available, then the train-free framework is a good way. Tip-Adapter is a strong baseline in the training-free methods, significantly improving the average by 33.6 percentage points without any parameter update. However, Tip-Adapter relies on the CLAP pre-trained model, which may not help our fiber acoustic data, as a significant domain shift exists. As shown in the table, \textbf{our proposed CLAP-S, which relies solely on external knowledge retrieval from the support set, achieves better results than Tip-Adapter}.

\noindent When training resources are available, fine-tuning the model can further improve the adaptation performance. \textbf{Our method, CLAP-S-$^+$, achieves the highest accuracy among all the baseline methods,} including Prompt Tuning, Adapter, Treff, and Tip-Adapter-F.  CLAP-S-$^+$ utilizes task-aligned embedding to repeated use of feature knowledge and thus further improves accuracy over the Tip-Adapter-F. Prompt Tuning performed poorly on our real-world dataset, with an accuracy of only 4-5\%. We hypothesize that this is due to Prompt Tuning being designed to maximize the use of pre-trained knowledge, which is less effective for tasks completely unseen by the pre-trained model.

\subsection{Few-Shot Adaptation}

\begin{figure}[htb!]
\centering
\vspace{-0.5em} 
\includegraphics[width=.45\textwidth]{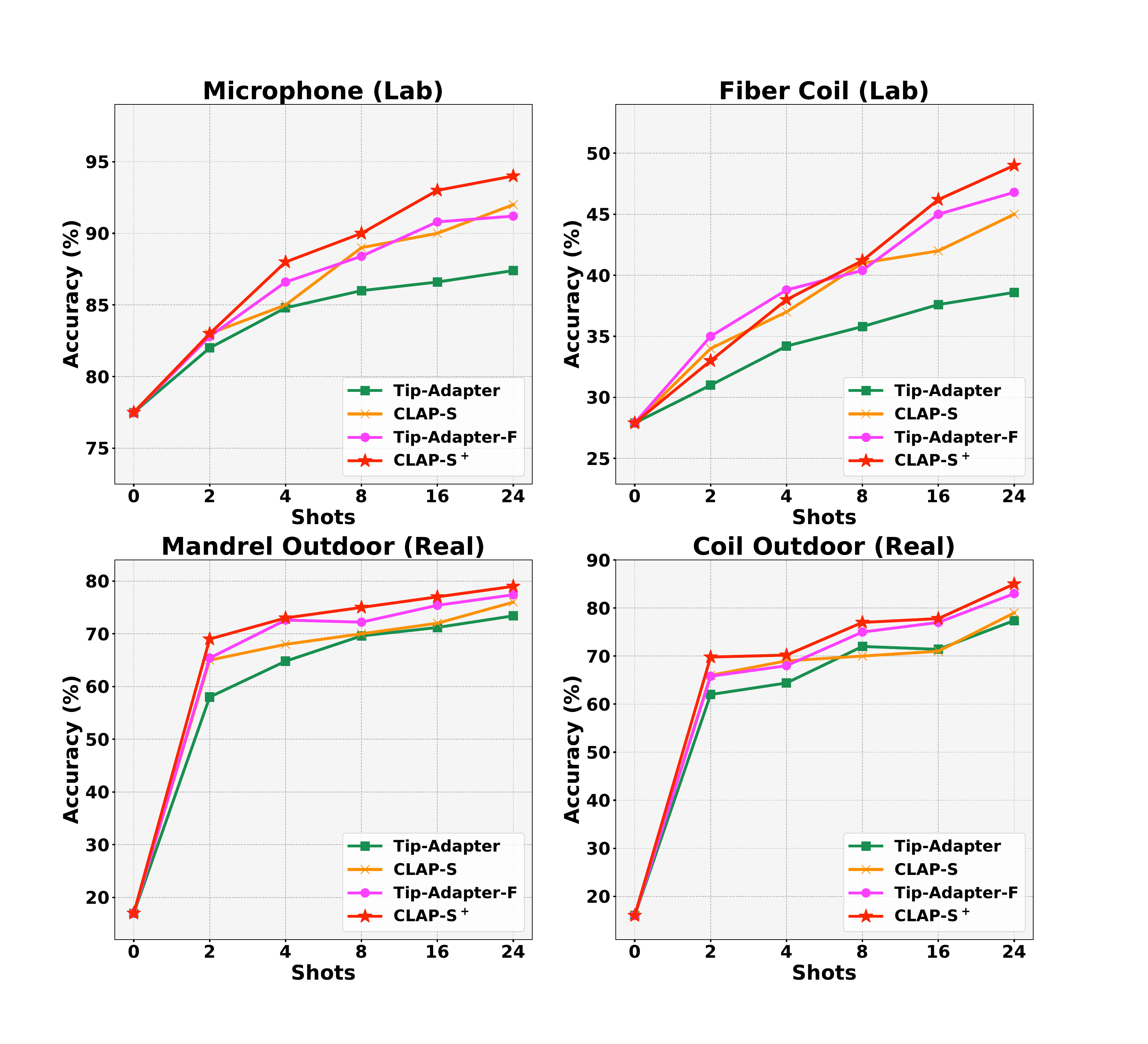}
\vspace{-1em} 
\caption{\textbf{The Few-Shot Adaptation Results.}}
\vspace{-1.5em} 
\label{few-shot}
\end{figure}

\noindent This scenario applies when we aim to minimize labeling cost and only limited data are available. We conducted few-shot experiments ranging from 2-shot to 24-shot, as shown in Figure \ref{few-shot}. The results demonstrate that both \textbf{CLAP-S and CLAP-S$^{+}$ consistently improve accuracy across the four datasets and perform competitively against baseline methods.} 

\begin{table}[]
\centering
\setlength{\tabcolsep}{4pt}
\fontsize{8}{9}\selectfont
\caption{Efficiency Comparison}
\begin{tabular}{@{}lccccc@{}}
\toprule
Method            & Acc    & Training & Parameters & Time  & Inference   \\ \midrule
Prompt Tuning     & 34.6    & Required & 6.14k      & 30min  & 266ms           \\
Tip-Adapter          & 69.2 & Free     & 0.0        & 5min   & 55ms            \\
\textbf{CLAP-S (ours) }         & 71.6 & Free     & 0.0        & 4min   & 45ms            \\
Adapter          & 76.4     & Required & 0.52M      & 15min  & 96ms            \\
Tip-Adapter-F & 76.5 & Required & 1.43M      & 10mins & 70ms            \\
\textbf{CLAP-S$^{+}$ (ours)}  & 78.8  & Required     & 0.52M      & 7mins  & 56ms            \\ \bottomrule
\end{tabular}
\label{table:efficiency}
\vspace{-1.5em}  
\end{table}

\subsection{Efficiency Comparison}
We also compared the efficiency of our method with other baselines, as shown in Table \ref{table:efficiency}. \textbf{CLAP-S demonstrated the best efficiency, requiring no training or additional parameter storage, with the lowest training and inference time.} However, this comes at the cost of sub-optimal performance. In contrast, our training-required version of CLAP-S$^{+}$ achieves the highest average accuracy across five datasets. 

\begin{table}[htb!]
\centering
\setlength{\tabcolsep}{1pt}
\fontsize{7.8}{8}\selectfont
\caption{Ablation Study }
\begin{tabular}{@{}lcccccc@{}}
\toprule
Methods            & ECM  & FM   & FC   & FMO   & FCO   & Avg \\ \midrule
\multirow{1}{*}{Support Set} &  92.0$_{\pm 0.1}$       & 61.4$_{\pm 0.8}$       & 43.0$_{\pm 0.0}$      & 79.0$_{\pm 0.1}$          & 83.0$_{\pm 0.1}$     & 71.6     \\\midrule
\multirow{1}{*}{ZS + Support Set} &  87.0$_{\pm 0.1}$       & 59.0$_{\pm 0.1}$       & 39.0$_{\pm 0.1}$      & 78.8$_{\pm 1.5}$          & 82.2$_{\pm 1.0}$     & 69.2    \\\midrule

\multirow{1}{*}{ZS + Support Set$^{+}$} & 91.8$_{\pm 1.0}$ & 58.6$_{\pm 1.0}$ & 39.2$_{\pm 1.8}$ & 25.0$_{\pm 0.1}$ & 20.0$_{\pm 0.1}$ & 46.9    \\\midrule
\multirow{1}{*}{Adapter} & 92.9$_{\pm 1.1}$        & 68.8$_{\pm 1.0}$        & 48.8$_{\pm 1.7}$       & 81.4$_{\pm 0.9}$           & 90.2$_{\pm 0.7}$    & 76.4     \\\midrule     

\multirow{1}{*}{Adapter + ZS} & 94.0$_{\pm 1.1}$        & 68.6$_{\pm 0.5}$        & 48.0$_{\pm 1.1}$       & 82.2$_{\pm 1.3}$           & 85.8$_{\pm 2.1}$    & 75.7     \\\midrule

\multirow{1}{*}{Adapter + Support Set}  & 93.4$_{\pm 1.4}$ & 71.6$_{\pm 1.0}$ & 50.6$_{\pm 1.0}$ & 84.0$_{\pm 1.5}$ & 88.3$_{\pm 1.3}$ & 77.6    \\\midrule
\multirow{1}{*}{Adapter + Support Set$^{+}$}  & \textbf{94.0}$_{\pm 1.2}$ & \textbf{70.0}$_{\pm 0.8}$ & \textbf{51.0}$_{\pm 1.2}$ & \textbf{87.0}$_{\pm 1.9}$ & \textbf{92.0}$_{\pm 1.7}$ & \textbf{78.8}    \\ 
\bottomrule
\end{tabular}
\label{ablation}
\vspace{-1.5em}  
\end{table}

\subsection{Ablation Study}
\noindent {\textbf{1: Does Zero-shot knowledge always contribute positively?}} 

The result in Table \ref{ablation} shows that by adding the zero-shot (`ZS' in the table) knowledge, the performance of the Support Set model and the Adapter model both drops.   Thus, \textbf{in our fiber acoustic domain, the zero-shot knowledge is not helpful}. This could be due to the misalignment of the text-audio representation caused by domain gaps and the deficiency of the language encoder in handling new acoustic concepts. This conclusion applies only to the specific domain considered, yet one can see that when significant domain gaps exist and acoustic events are difficult to describe with language, the zero-shot transfer may have unintended negative effects.


\noindent {\textbf{2: Which representation is more effective for retrieval in the support set — text-aligned or task-aligned?}}
 The text-aligned embedding for the Support Set is represented as `Support Set', while the task-aligned embedding for the Support Set is represented as `Support Set$^+$' in Table \ref{ablation}.  The results indicate that, for the same ZS or Adapter, adding `Support Set$^+$' consistently yields better performance than adding the `Support Set'. \textbf{Thus the task-aligned representation is more effective for key-value retrieval}. This is due to the larger domain gaps between the fiber-optic acoustic domain and the conventional microphone domain \cite{treff}. 


 \noindent {\textbf{3: Fine-tuning jointly or separately?}} In scenarios with data collected from multiple devices or neighboring channels, should we fine-tune one model or individual models separately? We conducted experiments and the results show \textbf{the model trained jointly on all the datasets outperformed the models trained independently} on each individual dataset (results shown in Table \ref{joint}). The reason is that joint training 
 serves as a form of data augmentation, where increased data improves performance on individual tasks.
 \vspace{-1em} 
 \begin{table}[htb!]
\centering
\setlength{\tabcolsep}{1pt}
\fontsize{8}{8}\selectfont
\caption{Jointly Training Vs Independently Training}
\begin{tabular}{@{}lcccccc@{}}
\toprule
Methods            & ECM  & FM   & FC   & FMO   & FCO   & Avg \\ \midrule
\multirow{1}{*}{Adapter-Independently} & 92.9$_{\pm 1.1}$        & 68.8$_{\pm 1.0}$        & 48.8$_{\pm 1.7}$       & 81.4$_{\pm 0.9}$           & 90.2$_{\pm 0.7}$    & 76.4     \\\midrule 
\multirow{1}{*}{Adapter-Jointly}  & \textbf{93.4}$_{\pm 0.9}$ & \textbf{70.8}$_{\pm 1.5}$ & \textbf{50.2}$_{\pm 1.6}$ & \textbf{81.9}$_{\pm 0.7}$ & \textbf{90.9}$_{\pm 0.5}$ & \textbf{77.4}    \\\bottomrule
\end{tabular}
\label{joint}
\vspace{-1.5em} 
\end{table}



\section{Conclusion}
\vspace{-0.5em} 
In this paper, we explored efficient model adaptation methods by balancing implicit knowledge and explicit knowledge for the challenging fiber-optic acoustic recognition task. Our work provides valuable insights into utilizing pre-trained models to design domain-shift mitigation strategies, and improve the robustness of text-aligned and task-aligned representations. Our method may also be applicable to other domains involving significant domain shifts from the pre-trained model (e.g., respiratory audio \cite{zhangtowards}). Extending our approach to other downstream tasks will be an interesting research direction.

\textbf{Acknowledgement }This work is partially supported by NSF AI Institute-2229873, NSF RI-2223292, an Amazon research award, and an Adobe gift fund. 



\bibliographystyle{IEEEbib}
\bibliography{refs}

\clearpage

\end{document}